\documentclass[twocolumn,showpacs,amsmath,nofootinbib,amssymb,epsfig]{revtex4}

\usepackage{graphicx}
\usepackage{dcolumn}
\usepackage{bm}

\begin{document}
\title{Stability of the Einstein static universe in Einstein-Cartan theory}
\author{K. Atazadeh }
\email{atazadeh@azaruniv.ac.ir}\affiliation{Department of Physics, Azarbaijan Shahid Madani University , Tabriz, 53714-161 Iran}

\date{\today}
\begin{abstract}
The existence and stability of the Einstein static solution have been built in the Einstein-Cartan gravity.
We show that this solution in the presence of perfect fluid with spin density satisfying the Weyssenhoff restriction is cyclically stable
around a center equilibrium point. Thus, study of this solution is interesting because it supports non-singular
emergent cosmological models in which  the early universe oscillates indeterminately
about an initial Einstein static solution and is thus past eternal.
\end{abstract}

\maketitle

\section{Introduction}
In 1923 \'{E}lie Cartan wrote a series of papers on
geometrical aspects of the theory of relativity. The papers contained
important new mathematical ideas which influenced the development
of differential geometry and, in particular, led to the general theory
of connections \cite{cartan1}. Cartan was remarkably, aware of the importance of the notion
of a connection and its relevance for physics. H. Weyl was
the first to introduce non-Riemannian linear connections by relaxing the condition of compatibility, $\nabla_{\nu}g_{\mu\nu}= 0$, between a metric
($g_{\mu\nu}$) and the connection. The generalization due to Cartan was more significant, he introduced connections with torsion
and a new law of parallel transport (the Cartan displacement).
Torsion and the Cartan displacement reflect the role played by the
group of affine transformations, an extension of the linear group to
translations. Attempting to treat torsion and curvature on the same footing, Cartan was led to a slight
modification of Einstein's relativistic theory of gravitation.
The modification, known today as the Einstein-Cartan theory (EC),
consists in relating torsion to the density of intrinsic angular momentum (spin) of matter, instead of assuming it to be zero, as done
in the Einstein theory \cite{cartan}.
The importance of the Cartan theory becomes more clear, if one tries to incorporate the
spinor field into the torsion-free general theory of relativity\cite{Wat}. In EC theory, the spin sources can be noted in terms of torsion, thus torsion is not a dynamical quantity \cite{Hehl}. Weyssenhoff exotic perfect fluid is one of the usual ways to
consider a fluid with intrinsic spin density \cite{Weys}.

In the context of alternative theories of gravity, cosmological solutions in the EC theory of gravity have been considered in which the spin properties of matter and their influence on the geometrical structure of space-time are studied. In Ref.\cite{Kuch}, the effects of torsion and spinning matter in cosmological context such as inflationary scenarios, late time acceleration of the universe and removing the singularities are investigated. Recently, in \cite{sharam} the authors have shown that the spin-spin contact
interaction in the EC gravity can lead to signature changing solutions.

In the framework of Einstein's general relativity (GR), a new scenario, called an emergent universe was introduced in  \cite{3,4} to remove initial singularity. Observation from WMAP7 \cite{5} supports the positivity of space curvature in which it is found that a closed universe is favored at the 68 \%\ confidence level, and the universe stays, past-eternally,
in an Einstein static state and then evolves to a subsequent inflationary phase.
According to emergent theory of gravitation, the universe might have been originated from an Einstein static state rather
than a big bang singularity. Nevertheless, the Einstein static universe in the Einstein's
general relativity is unstable, which means that it is almost impossible for the universe to
maintain its stability  in a long time because of the existence of varieties of perturbations,
such as the quantum fluctuations. Thus, it seems to us that the original emergent model
dose not resolve the big bang singularity problem properly as expected.
However, in the early epoch, the universe is apparently under fanatical physical conditions,
the study of the initial state may be affected by novel physical effects, such
as those resulting from quantum gravity, or a modified gravity or
even other new physics.
In \cite{Boehmer}, the author has considered the sign problem of the cosmological constant in the context of EC theory for a
static and spherically symmetric  Einstein static universe.
Finally,  the stability of the Einstein static state has been studied
in various cases \cite{Carneiro, Mulryne, Parisi, Wu,
Lidsey, Bohmer2007, Seahra, Bohmer2009, Barrow,
Barrow2009, Clifton,Boehmer2010, Boehmer20093, Wu20092,
Odrzywolek}, from loop quantum gravity \cite{Mulryne, Parisi, Wu} to $f(R)$ gravity  \cite{Boehmer2010} and $f(T)$ gravity \cite{ft}, from Horava-Lifshitz gravity \cite{Wu20092,Odrzywolek} to brane gravity \cite{Lidsey} and massive gravity \cite{mass}.

In this paper, we consider the stability of the Einstein static universe in the Friedmann-Lema\^{\i}tre-Robertson-Walker space-time in the
framework of EC gravity with exotic Weyssenhof perfect fluid.  In Section III we present an analysis of the
equilibrium of Einstein solution in the presence of matter and the spin. Next, we consider a numerical example,
in which the energy contain relativistic matter and a spin fluid with negative energy and negative pressure.
The paper ends with a brief conclusions in Section IV.

\section{Friedmann equation in spin-Dominated Einstein Cartan gravity}

Einstein-Cartan gravity can be started by writing the following action
\begin{eqnarray}\label{1}
{\cal S}=\int\sqrt{-g}\left[-\frac{1}{16\pi G}\left(\tilde
R-2\Lambda\right)+{\cal L}_M\right]d^4x,
\end{eqnarray}
where $\tilde R$ is the Ricci scalar associated to the asymmetric
connection $\tilde\Gamma^\mu_{\,\,\,\,\rho\sigma}$ and $\Lambda$ is
the cosmological constant, also ${\cal L_M}$ is the Lagrangian density of matter
fields. Utilizing of the metric compatibility
$\tilde\nabla_\rho g_{\mu\nu}=0$ \cite{Hehl} and the
definition of torsion $T^\mu_{\,\,\,\,\rho\sigma}=\tilde\Gamma^\mu_{\,\,\,\,\rho\sigma}-\tilde\Gamma^\mu_{\,\,\,\,\rho\sigma}$,
the connection $\tilde\Gamma^\mu_{\,\,\,\,\rho\sigma}$ can be
written as
\begin{eqnarray}\label{2}
\tilde\Gamma^\mu_{\,\,\,\,\rho\sigma}=\Gamma^\mu_{\hspace{.2cm}\rho\sigma}+K^\mu_{\hspace{.2cm}\rho\sigma},
\end{eqnarray}
where $\Gamma^\mu_{\,\rho\sigma}$ and $K^\mu_{\hspace{.2cm}\rho\sigma}$ are the Christoffel symbol and the
contorsion tensor, respectively, which are related to the torsion
$Q_{\rho\sigma}^{\hspace{.5cm}\mu}=\tilde
\Gamma_{[\rho\sigma]}^{\hspace{.5cm}\mu}$ via
\begin{eqnarray}\label{3}
K^\mu_{\hspace{.2cm}\rho\sigma}
=\frac{1}{2}\left(Q_{\hspace{.2cm}\rho\sigma}^{\mu}-Q^{\hspace{.3cm}\mu}_{\rho\hspace{.4cm}\sigma}-
Q^{\hspace{.3cm}\mu}_{\sigma\hspace{.4cm}\rho}\right).
\end{eqnarray}

By variation of the action with respect to the metric and contorsion,
one can find the equations of motion \cite{Hehl}
\begin{eqnarray}\label{4}
G^{\mu\nu}-\Lambda
g^{\mu\nu}-\left(\nabla_{\rho}+2Q_{\rho\sigma}^{\hspace{.4cm}\sigma}\right)\times \\\nonumber\left(T^{\mu\nu\rho}-T^{\nu\rho\mu}+T^{\rho\mu\nu}\right)=8\pi
GT^{\mu\nu},\\\nonumber
T^{\mu\nu\rho}=8\pi G\tau^{\mu\nu\rho},
\end{eqnarray}
where
\begin{eqnarray}\label{5}
T_{\mu\nu}^{\hspace{.3cm}\rho}=Q_{\mu\nu}^{\hspace{.3cm}\rho}+\delta_\mu^\rho
Q_{\nu\sigma}^{\hspace{.4cm}\sigma}-\delta_\nu^\rho
Q_{\mu\sigma}^{\hspace{.4cm}\sigma},
\end{eqnarray}and $G^{\mu\nu}$ and $\nabla_\rho$ are the usual Einstein tensor and covariant derivative for the full
nonsymmetric connection $\Gamma$, respectively. Also, the canonical spin-density and the energy-momentum tensors are given by
\begin{eqnarray}\label{6}
\tau^{\mu\nu\rho}=\frac{1}{\sqrt{-g}}\frac{\delta{\cal
L}_M}{\delta K_{\rho\nu\mu}}~~~~~~~~~
T^{\mu\nu}=\frac{2}{\sqrt{-g}}\frac{\delta{\cal L}_M}{\delta g_{\mu\nu}},
\end{eqnarray}
respectively. Thus, by means of equations (\ref{4}) and (\ref{5}) one can
write generalized Einstein field equations as
\begin{eqnarray}\label{7}
G^{\mu\nu}(\Gamma)=8\pi G(T^{\mu\nu}+\tau^{\mu\nu}),
\end{eqnarray}
where $G^{\mu\nu}(\Gamma)$ is the known symmetric Einstein tensor
and
\begin{eqnarray}\label{8}
\tau^{\rho\sigma}&=&
\left[-4\tau^{\rho\mu}_{\hspace{.4cm}[\nu}\tau^{\sigma\nu}_
{\hspace{.3cm}\mu]}-2\tau^{\rho\mu\nu}\tau^\sigma_{\hspace{.3cm}\mu\nu}+\tau^{\mu\nu\rho}\tau_{\mu\nu}^{\hspace{.3cm}\sigma}+
\right.\nonumber\\&&\left.\frac{1}{2}g^{\rho\sigma}
\left(4\tau_{\lambda
\hspace{.1cm}[\nu}^{\hspace{.2cm}\mu}\tau^{\lambda\nu}_{\hspace{.4cm}\mu]}+\tau^{\mu\nu\lambda}\tau_{\mu\nu\lambda}\right)\right],
\end{eqnarray}
is a kind of modification to the space-time curvature that stems from the spin
\cite{Weys}. If we set the spin zero in equation (\ref{7}) we will have  the standard Einstein field equations.
We suppose that ${\cal L}_M$ represents a fluid of spinning particles in the early Universe
minimally coupled to the metric and the torsion of the $U_4$ theory.
In the case of the spin fluid the canonical spin tensor is given by \cite{Weys}
\begin{eqnarray}\label{9}
\tau^{\mu\nu\rho}=\frac{1}{2}S^{\mu\nu}u^\rho,
\end{eqnarray}
where $u^\rho$ is the 4-velocity of the fluid and $S^{\mu\nu}$ is the antisymmetric spin density \cite{Ob}. Then, the energy-momentum
tensor can be separated into the two parts: the usual perfect fluid
$T_F^{\hspace{.1cm}\rho\sigma}$ and an intrinsic-spin part
$T_S^{\hspace{.1cm}\rho\sigma}$, as
\begin{equation}
T^{\rho\sigma}
= T_F^{\hspace{.1cm}\rho\sigma}+
T_S^{\hspace{.1cm}\rho\sigma},
\end{equation}
thus, the explicit form of intrinsic-spin part is given by
\begin{eqnarray}\label{10}
&&T_S^{\hspace{.2cm}\rho\sigma}=u^{(\rho}S^{\sigma)\mu}u^\nu
u_{\mu;\nu}+(u^{(\rho}S^{\sigma)\mu})_{;\mu}+\\\nonumber&&Q_{\mu\nu}^{\hspace{.2cm}(\rho}u^{\sigma)}
S^{\nu\mu}-u^\nu
S^{\mu(\sigma}Q^{\rho)}_{\hspace{.1cm}\mu\nu}-\omega^{\mu(\rho}S^{\sigma)}_{\hspace{.3cm}\mu}+u^{(\rho}S^{\sigma)\mu}\omega_{\mu\nu}u^\nu,
\end{eqnarray}
where $\omega$ and semicolon denote the angular velocity associated with the intrinsic
spin and covariant derivative with respect to
Levi-Civita connection, respectively. According to the usual explanation of EC gravity we can
assume that $S_{\mu\nu}$ is associated with the quantum mechanical
spin of microscopic particles \cite{Kuch}, thus for unpolarized
spinning field we have $<S_{\mu\nu}>=0$ and if we consider
\begin{equation}
\sigma^2=\frac{1}{2}<S_{\mu\nu}S^{\mu\nu}>,
\end{equation}
we get
\begin{eqnarray}\label{11}
<\tau^{\rho\sigma}>=4\pi G \sigma^2u^\rho u^\sigma +2\pi G
\sigma^2g^{\rho\sigma},
\end{eqnarray}
and
\begin{eqnarray}\label{12}
<T_F^{\hspace{.1cm}\rho\sigma}>=(\rho+p)u^\rho u^\sigma-pg^{\rho\sigma}\\\nonumber
<T_S^{\hspace{.2cm}\rho\sigma}>=-8\pi G\sigma^2u^\rho u^\sigma.
\end{eqnarray}
This leads to the simplification of EC generalization of standard gravity as follow
\begin{eqnarray}\label{13}
G^{\rho\sigma}(\Gamma)=8\pi G \Theta^{\rho\sigma},
\end{eqnarray}
where $\Theta^{\rho\sigma}$ explains the effective macroscopic
limit of matter field
\begin{eqnarray}\label{14}
\Theta^{\rho\sigma}&=&<T^{\rho\sigma}>+<\tau^{\rho\sigma}>\\\nonumber&=&\left(\rho+p-\rho_{s}-p_{s}\right)u^\rho u^\sigma-\left(p-p_{s}\right)g^{\rho\sigma},
\end{eqnarray}
where $\rho_s=2\pi G \sigma^2$.
In comparison with the usual GR, we can conclude that equations
(\ref{13}) and (\ref{14}) show the equality between EC field equations and
the Einstein equations coupled to a fluid with a
particular equation of state as the matter source. Actually, in a
hydrodynamical concept the contribution of the torsion can be
done by a spin fluid such that
\begin{equation}\label{rev1}
\rho_{tot}=\rho-2\pi G \sigma^2,\hspace{5mm} p_{tot}=p-2\pi G \sigma^2.
\end{equation}
It is considerable to recall that the correction
terms signs in (\ref{rev1}) are  compatible with the
semi-classical models of spin fluid which are negative \cite{Weys}, \cite{Kuch}. In other words  the effect of spin in EC theory plays the role of  a perfect fluid with negative energy density and pressure. In such a model the Einstein static universe
occurs and it is stable around equilibrium point.\\
Inserting closed isotropic and homogeneous Friedmann-Lema\^{\i}tre-Robertson-Walker line element into the
(\ref{13}) and (\ref{14}) gives the field equations
\begin{eqnarray}\label{17}
3H^2+\frac{3}{a^2}=8\pi G(\rho-\rho_s),
\end{eqnarray}
where $H=\frac{\dot{a}}{a}$ is the Hubble parameter. The conservation equation gives
\begin{eqnarray}\label{18}
\frac{d}{dt}(\rho-\rho_s)=-3H(\rho+p-p_s-\rho_s).
\end{eqnarray}
Equation (\ref{18}) is a generalized form of the covariant energy conservation law
by including the spin.
To continue, we take the matter field as a
unpolarized fermionic perfect fluid with equation of state
$p=w\rho$. Finally, we have
\begin{equation}
\sigma^2=\frac{1}{2}<S^2>=\frac{1}{8}\hbar^2<n^2>,
\end{equation}
where $n$ is the particle number density, and averaging
process gives \cite{Nurgaliev}
\begin{eqnarray}\label{19}
\sigma^2=\frac{\hbar^2}{8}B_w^{-\frac{2}{1+w}}\rho^{\frac{2}{1+w}},
\end{eqnarray}
where $B_w$ is a dimensional constant dependent on $w$.
Thus, from the conservation equation (\ref{18}) we can write
\begin{eqnarray}\label{20}
\rho=\rho_0a^{-3(1+w)},
\end{eqnarray}
where $\rho_0$ is present value of energy density. For
simplicity, we define
\begin{equation}
D=\frac{8\pi G}{32}\hbar
B_w^{-\frac{2}{1+w}}\rho_0^{\frac{2}{1+w}},
\end{equation}
therefore, from equations (\ref{rev1}), (\ref{19}) and (\ref{20}) $\rho_s $ can be written as
\begin{eqnarray}\label{21}
\rho_s=Da^{-6},
\end{eqnarray}
Note that effects of spin are dynamically equivalent to introducing into the model some additional
non-interacting fluid for which the equation of state is $p_s=w_{s}\rho_s$ where $w_s=1$, $\rho_s\propto a^{-6}$, denotes for stiff matter  or brane effects with dust on a brane with negative tension.

\section{The Einstein static solution and stability}

By using equations (\ref{17}) and (\ref{18}) the Raychadhuri equation can be written as\footnote{We have set units $8\pi G=1$.}
\begin{eqnarray}\label{aa}
\ddot{a}=-\frac{\dot{a}^{2}+1}{2a}(1+3w)+\frac{D}{2a^{5}}(w_{s}-w).
\end{eqnarray}
The Einstein static solution is given by $\ddot{a} = 0 = \dot{a}$. To begin with we obtain the conditions
for the existence of this solution. The scale factor in this case is given by
\begin{eqnarray}
a^{4}_{_{Es}}=\frac{D(w_{s}-w)}{1+3w}.
\end{eqnarray}
The existence condition reduces to the reality condition for $a_{_{Es}}$, which for a positive $D$ takes
the forms
\begin{eqnarray}\label{ww}
w>-1/3 ~~~~{\rm and} ~~~w<w_{s},
\end{eqnarray}
or
\begin{eqnarray}\label{www}
w<-1/3 ~~~~{\rm and} ~~~w>w_{s}.
\end{eqnarray}

Here, we are going to study the stability of the critical point. For convenience,
we introduce two  variables
\begin{eqnarray}
x_1=a,\quad x_2=\dot{a}.
\end{eqnarray}
It is then easy to obtain the following equations
\begin{eqnarray}
\dot{x}_1=x_2,
\end{eqnarray}
\begin{eqnarray}
\dot{x}_2=-\frac{{x}_2^{2}+1}{2x_1}(1+3w)+\frac{D}{2x_1^{5}}(w_{s}-w).
\end{eqnarray}
According to these variables, the fixed point, $x_1=a_{Es},\; x_2=0$ describes the Einstein static solution properly. The stability of the critical
point is determined by the eigenvalue of the coefficient matrix ($J_{ij}=\frac{\partial \dot{x}_{i}}{\partial x_{j}} $)
stemming from linearizing the system explained in details by above two equations near the critical point. Using $\lambda^2$ to obtain the
eigenvalue we have
\begin{eqnarray}
\lambda^2=\frac{-2D}{a_{Es}^{6}}(w_{s}-w).
\end{eqnarray}
In the case of $\lambda^{2}< 0$ the Einstein static solution has a center
equilibrium point, so it has circular stability, which means that small perturbation from the fixed
point results in oscillations about that point rather than exponential deviation from it. In
this case, the universe oscillates in the neighborhood of the Einstein static solution
indefinitely.
Thus, the stability condition is determined by $\lambda^{2}< 0$. For $D> 0$, this means
that $w<w_{s}$. Comparing this inequality with the conditions for existence of the Einstein
static solution, (\ref{ww}) and (\ref{www}), we find that the Einstein universe is stable $w>-1/3$.
Especially, it is stable in the presence of ordinary matter ($ w$) plus a spin fluid with negative energy density and negative pressure.
\begin{figure*}[ht]
  \centering
  \includegraphics[width=3in]{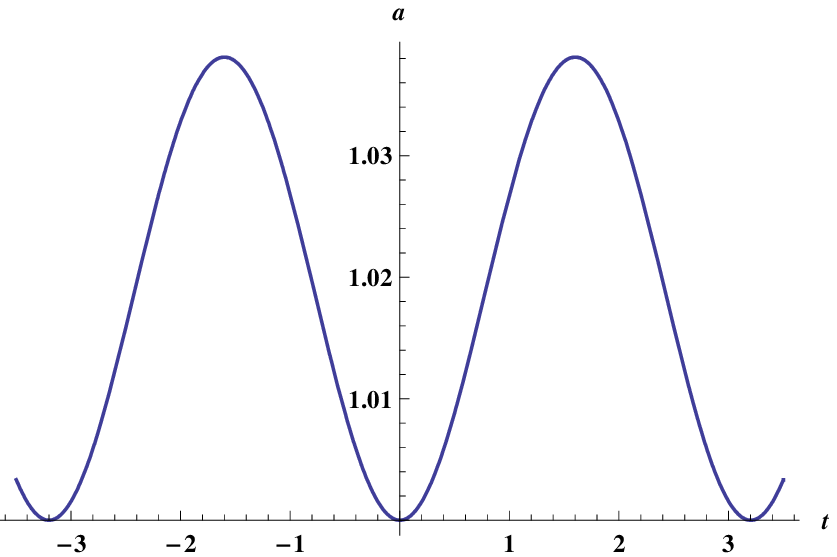}~~~~~~~~~~~~~~~
   \includegraphics[width=3in]{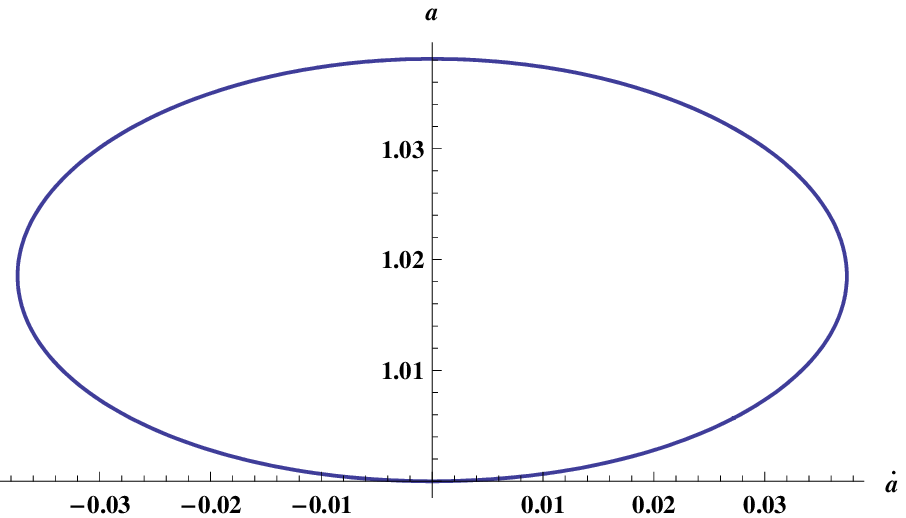}
  \caption{The evolutionary curve of the scale factor with time (left) and
the phase diagram  in space ($a$, $\dot{a}$) (right) for $w = 1/3$.}
  \label{stable}
\end{figure*}

To continue, we study the effects of spin field in EC
on the dynamics of the universe. As an example, we consider the case where the energy
content consists of spin fluid, which we put $w_s = 1$ and $\rho_s = D/a^{6}$, in addition to
a relativistic matter with an equation-of-state parameter $w = 1/3$. Using these equation of
state parameters in equation (\ref{aa}) we obtain
\begin{eqnarray}
3a^{5}\ddot{a}+3a^{4}\dot{a}^{2}+3a^{4}-D=0
\end{eqnarray}

From the above equation the corresponding scale factor of Einstein static solution is given by $a^{4} = D/3$.
Obviously, phase space trajectories which is beginning precisely on the Einstein static fixed point remain there indeterminately. From another point of view, trajectories which are creating in the vicinity
of this point would oscillate indefinitely near this solution.
An example of such a universe trajectory using initial conditions given by $a(0) = 1$ and $\dot{a}(0) = 0$,
with $D = 3.23$ has been plotted in Fig. 1.

\section{CONCLUSION}

We have discussed the existence and stability of the Einstein static universe in the presence
of spin fields coupled to gravity through the EC gravity. We have shown that the
spin energy density in Einstein universe is proportional to the inverse sixth power of
the scale factor. Also, we have determined the allowed intervals for the equation of state parameters related to the spin energy
such that the Einstein universe is stable, while it is dynamically belonging
to a center equilibrium point. The motivation study of such a solution is the result of its essential role in
the construction of non-singular emergent oscillatory models which are past eternal, and hence can resolve
the singularity problem in the standard cosmological scenario.
\section*{Acknowledgments}
The author is grateful to F. Darabi for useful discussion and recommendation.

\end{document}